\documentclass[letter,twocolumn]{jpsj2}
\usepackage{amsmath,amssymb,graphicx,bm}
\newcommand{\eq}{eq. }
\newcommand{\eqs}{eqs. }
\newcommand{\fig}{Fig. }
\title{Electron transport driven by a chemical potential difference}
\author{Chihiro \textsc{Nakajma}$^{1}$\thanks{E-mail: nakajima@yukawa.kyoto-u.ac.jp}
and Hisao \textsc{hayakawa}$^{1}$\thanks{E-mail: hisao@yukawa.kyoto-u.ac.jp}}

\inst{$^{1}$Yukawa Institute for Theoretical Physics, Kyoto University, \\
Kitashiraoiwake cho, Sakyo-ku, Kyoto 606-8502}

\abst{
Based on Bhatnagar-Gross-Krook  equation coupled with Maxwell equation,
we investigate the spatial dependence of a chemical, an electrostatic and an electrochmeical potentials 
inside a specimen connected with  reservoirs. 
We also confirm that a gap of the chemical potential between at a connection point is negligible.
}
\kword{electron transport, BGK equation, chemical potential}

\begin{document}
\maketitle
Study of the electron transport has  a long history \cite{sommerfeld} and 
is one of the typical problems in the nonequilibrium statistical physics. 
In these days 
there has been an increasing interest in understanding the electron transport 
by the successful fabricated devices  \cite{WF}.
In the nonequilbrium transport theory, 
it is noted that electron currents and heat currents are related with 
the temperature gradient, the chemical potential gradient and the electric field \cite{ziman2}.
For these thermoelectric relations,   Wiedemann-Franz law \cite{WF} and Seebeck effect \cite{seebeck}
are well known.
On the other hand, it is interesting to investigate 
how the local thermodynamical quantities are distributed in the conductance.
In this leter, we focus on the spatial dependence of the thermodynamical quantities 
in the steady electric current which is driven by the chemical potential difference.

\begin{figure}
\begin{center}
\includegraphics[width=60mm]{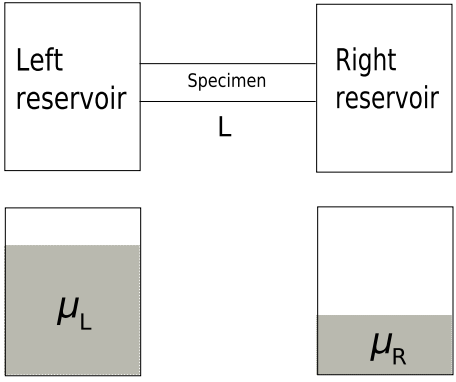} \\ 
\caption{The schematic picture of  the electron transport: 
The specimen is directly connected with two equilibrium reservoirs.
The reservoirs have chemical potential $\mu_{L}$ and $\mu_{R}$ respectively.
The length of the specimen is $L$.}\label{model}
\end{center} 
\end{figure}

Let us consider the electronic device which is composed of a specimen connected through 
reservoirs.
The electric current is driven by the difference of the chemical potential in both sides of the reservoirs.
We assume that the system is  at the isothermal low temperature
and  any effects of  the heat current do not exist.
Furthermore, Joule heat caused by the electric current assumed to be negligible. 
At the connection point between the specimen and the reservoir, 
we need to be careful about the discontinuity of  the thermodynamical quantities
 \cite{michael,nishino}.
The schematic picture of the device is given by \fig \ref{model}.
The length of the specimen is larger than the inelastic scattering length and 
its specimen is directly connected with two equilibrium reservoirs.
Here, we do not consider the collimator effect at the connection point\cite{payne,shimizu}.
Each reservoir  has the chemical potentials $\mu_{L}$  and  $\mu_{R}$ 
which keep  constant values during the steady electric current. 

Although the interactions between the electrons play an important  role for 
strongly correlated electron systems, 
we can use the nearly free electron model to  explain
the  electric current in the metal.
In this letter, we treat the electrons as semi-classical particles which are
represented by a nonequilibrium distribution function.
To describe such electron transports, we adopt  BGK (Bhatnagar-Gross-Krook)  equation  \cite{BGK}  which is 
Boltzmann equation with a relaxation time approximation.
Furthermore, we couple  BGK equation with classical Maxwell equation.
Under this setup,  we investigate the spatial dependence of the chemical, electrostaic and electrochemical potentials  in the specimen,
and the discontinuity of the chemical potential at the connection point.

We start with BGK equation to treat the electron transport.
Here, we are interested in the $x$-directional spatial dependence of the thermodynamical quantities,
so that we consider the averaged behavior of the electric current in the cross section.
The steady-state BGK equation for the  distribution function $f(\bm{p})$ is 
\begin{equation}
 v_{x} \frac{\partial f}{\partial x}
-e E \frac{\partial f}{\partial p_{x}}
=I[f],
  \label{eq:boltzmann}
\end{equation}
where $-e$ is the charge of the electron, $E$ is the $x$-component of  the electric field, 
$p_{x}$ is the $x$-component of the linear momentum and the collison term 
$I[f]$ is given by
\begin{equation}
I[f]=-\frac{f -f^{0}}{\tau}, \label{tauf} 
\end{equation}
where $\tau$ is the relaxation time and
$f^{0}=1/(e^{\beta(\epsilon(\bm{p})-\mu)}+1)$
is Fermi distribution function with the equilibrium chemical potential $\mu$,
and $\epsilon(\bm{p})=\bm{p}^2/2m$ is the kinetic energy of the electron. 
The electric field has a contribution from the chemical potential difference and an 
internal contribution from the induced charge density fluctuations.
Here we do not consider the magnetic field induced by electric current.
When the distribution function $f$ is slightly different from 
the equilibrium Fermi distribution function \cite{ziman2},  
we obtain the solution
 \begin{equation}
   \label{eq:shift}
   f=f^{0} - f^{0}(1-f^{0}) e\beta\tau v_{x}\biggl(E+\frac{1}{e}\frac{d\mu}{dx}\biggr).
 \end{equation}
The most relevant processes to determine the relaxation time $\tau$ in electric conduction at low temperature
are electron-impurity scatterings. 
Employing Born approximation, we can evaluate $\tau$ by the chemical potential $\mu$ in the low temperature limit:
\begin{equation}
\label{tau}
  \tau^{-1}=\frac{16\sqrt{2}\pi^{3}m^{3/2}}{h^4}u_{0}^{2}n_{\text{imp}}\mu^{1/2},
\end{equation}
where $n_{\text{imp}}=N_{\text{imp}}/V$ is the impurity density, $m$ is the electron mass, $h$ is Planck constant
and $u_0$ is connected to the scattering potential $V(\bm{r})$ as $V(\bm{r})=u_0\sum_i\delta(\bm{r}-\bm{R}_i)$ with
the position of the random impurity $\bm{R}_i$.\cite{abrikosov} 

From \eqs  \eqref{eq:shift} and \eqref{tau}, the current density is given by
 \begin{equation}
 j=-2e\int \frac{d\bm{p}}{h^3} v_{x}f,\label{jj}
\end{equation}
where factor 2 is the spin degeneracy.
Furthermore, 
we have to take into account the charge imbalances that appear in the specimen.
The electric field arises due to the local deviation of the electron density
from its equilibrium value and satisfies the Maxwell equation

\begin{equation}
    \frac{dE}{dx}=-\frac{e}{\epsilon}
                   [2\int\frac{d\bm{p}}{h^3}f -\rho(x)],
\label{E}
\end{equation}
where $\epsilon$ is the  permittivity and
the unperturved electron density $\rho(x)$
is
\begin{equation}
    \rho(x)=2\int \frac{d\bm{p}}{h^3}f^{0}.
\label{rho}
\end{equation}

Now, we nondimensionalize  the physical quantities (atomic units) as 
$j \gets j\hbar^7(4\pi\epsilon_{0})^4/m^3 e^9$,
$v_{x}\gets v_{x}\hbar 4\pi\epsilon_{0}/e^2$,
$x\gets xm e^2/\hbar^2 4\pi\epsilon_{0}$,
$\mu\gets \mu\hbar^2 (4\pi\epsilon_{0})^2/m e^4$,
$E\gets E \hbar^4 (4\pi\epsilon_{0})^3/m^2 e^5$,
$n_{imp}\gets n_{imp}\hbar^6 (4\pi\epsilon_{0})^3/m^3 e^6$,
$u_{0}\gets u_{0}m^2 e^2/\hbar^4 4\pi\epsilon_{0}$.
Applying Sommerfeld expansion \cite{landau} to \eq \eqref{jj},
the dimensionless steady current density  is given by
\begin{align}
j= \frac{2}{3 \pi} \frac{1}{u_0^2 n_{\text{imp}}}
\mu(E+\frac{d \mu}{d x}),\label{jk3}
\end{align}
where we notice that $j$ is the negative value.
Similarly, from \eqs \eqref{E} and  \eqref{rho}, dimensionless Maxwell equation becomes
\begin{equation}
    \label{EE}    
\frac{dE}{dx}=\frac{8}{\sqrt{6}}\frac{1}{\epsilon_{r}u_{0}^{2}n_{imp}}
\sqrt{\mu}(E+\frac{d\mu}{dx}),
\end{equation}
with the relative permittivity $\epsilon_{r}=\epsilon/\epsilon_{0}$.

\begin{figure}
\begin{center}
    \includegraphics[width=95mm]{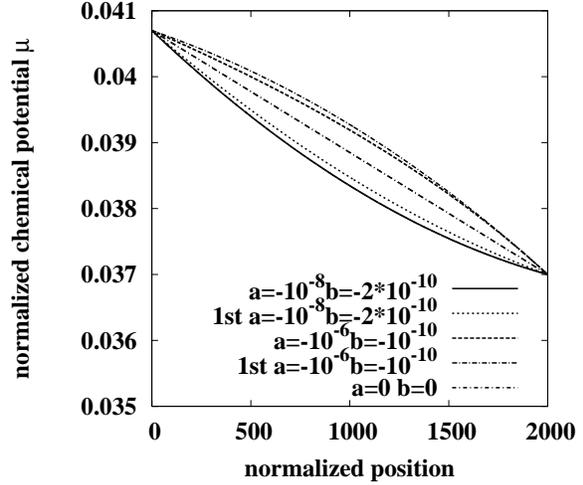} \\ 
\caption{Chemical potential profiles for the coefficient 
$(a,b)=(-10^{-8},-2\times10^{-10})$ ,
$(-10^{-6},-10^{-10})$ 
and $(0,0)$. 
Line 1st are  the chemical potential profile for the first order perturbative solution
with the corresponding coefficients.
The ratio of the chemical potential $\mu_{L}/\mu_{R}$ is 1.1. 
The length of the specimen is L=2000.
}\label{L}
\end{center} 
\end{figure}
\begin{figure}
\begin{center}
    \includegraphics[width=95mm]{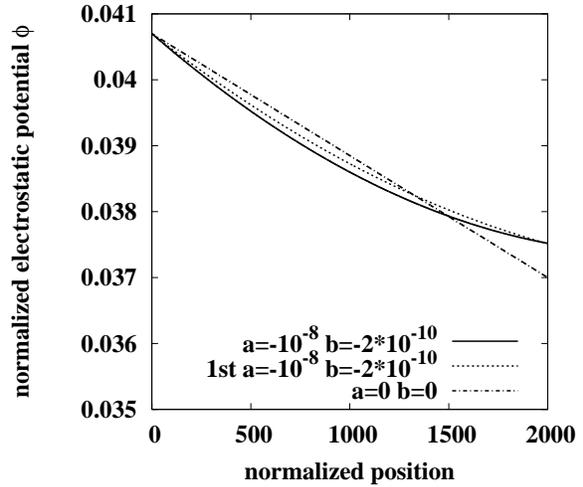} \\ 
    \includegraphics[width=95mm]{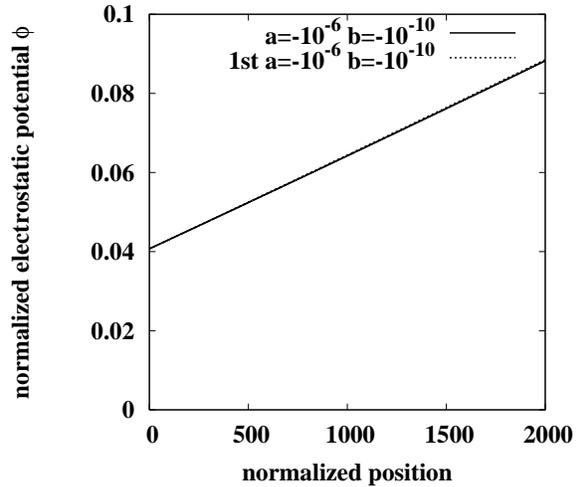} \\ 
\caption{Top(A): Electrostatic potential profile for the coefficient 
 (a,b)=$(-10^{-8},-2\times10^{-10})$ and (0,0).
Bottom(B): Electrostatic potential profile for the coefficient $(-10^{-6},-10^{-10})$.
Line 1st are  the electrostatic potential profile for the first order perturbative solution  in \eq \eqref{phii}. 
All the remaining parameters are the same as in Fig. 2. }\label{P}
\end{center} 
\end{figure}

Let us consider the chemical potential inside the specimen.
From \eqs \eqref{jk3} and \eqref{EE},
we obtain the ordinary differential equation of the chmeical potential:

\begin{equation}
    \frac{d^2 \mu}{dx^2}+a\frac{1}{\mu^2}\frac{d\mu}{dx}+b\mu^{-1/2}=0,
\label{mu}
\end{equation}
where  dimensionless coefficients $a$ and $b$ are given by
\begin{equation}
    \begin{cases}
        &a=\frac{3\pi}{2}u^{2}_{0}n_{imp}j\\
        &b=\frac{12\pi}{\sqrt{6}\epsilon_{r}}j.\label{ab}
    \end{cases}
\end{equation}
We now solve \eq \eqref{mu} subject to the boundary conditions 
at $\mu(0)=\mu_{L}$ and $\mu(L)=\mu_{R}$.
If we know the details of the material properties, 
the amount of the current density under the chemical potential difference
can be determined.
However, since the impurity potential $u_{0}$ is hard to be measured in the experiment,
we do not know the amount of the current density precisely.
In this letter,
we determine the spatial dependence of the chemical potential
as a function of $a$ and $b$.

Let us set the ratio $\mu_{L}/\mu_{R}$=1.1 with its normalized chemical potential 
$\mu_{L}$=0.0407, $\mu_{R}$=0.0370 and the normalized length 
$L$=2000
($\mu_{L}$=1.1eV, $\mu_{R}$=1.0eV, $L$=106nm).
Figure. \ref{L} shows the spatial dependence of the chemical potental
for the different coefficients $a$ and $b$. 
In the finite current density, the chemical potential profiles exhibit the strong nonlinearity.
When the coefficient $(a,b)$ is $(-10^{-8},-2\times10^{-10})$,
the spatial variation of the chemical potential has a concave profile,
while the chemical potential profile with the coefficient $(-10^{-6},-10^{-10})$  shows a convex one.
There exists the solutions of \eq \eqref{mu}
only when the value of $b$ is smaller than $10^{-6}$ for any coefficients $a$.

The spatial dependence of the electrostatic potential in the specimen
is of interest.
The electrostatic potential is related to the electric field:
$\phi=-\int E dx$.
Using the electric field $E=a\mu^{-1}- d\mu/dx$ from \eq \eqref{jk3},
the spatial dependence of the electrostatic potential can be determined. 
In \fig \ref{P}(A), the electrostatic potential profile with the 
coefficient $(-10^{-8},-2\times10^{-10})$ 
shows a concave one.
On the other hand, in \fig \ref{P}(B), 
the electrostatic potential profile with the coefficient $(-10^{-6},-10^{-10})$ almost increases linearly
and shows a larger spatial variation compared with the corresponding chemical potential profile  in \fig \ref{L}.
\begin{figure}
\begin{center}
    \includegraphics[width=95mm]{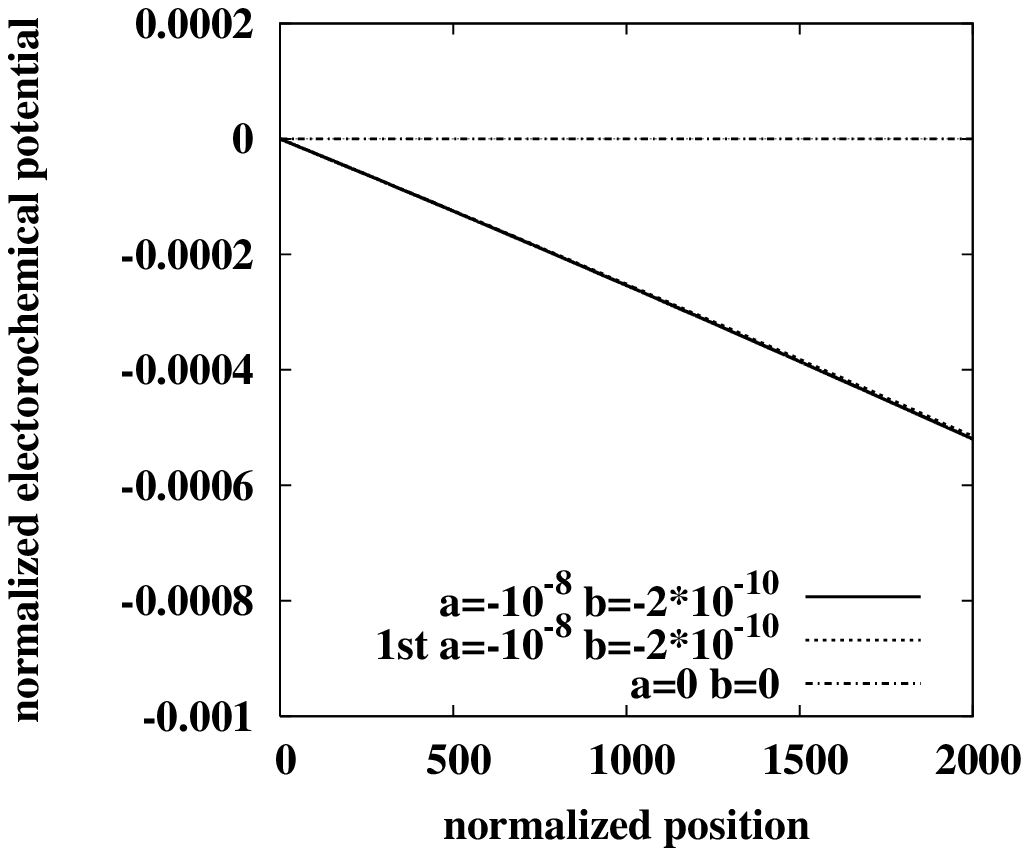} \\ 
    \includegraphics[width=95mm]{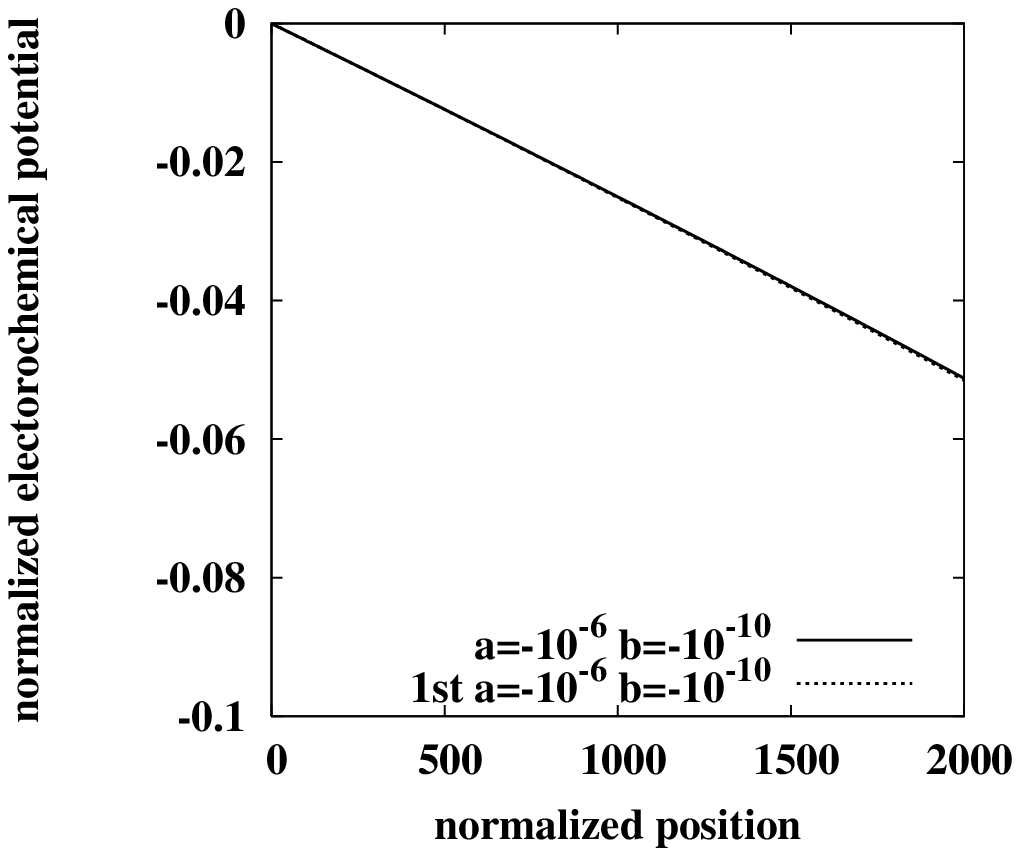} \\ 
  \caption{Top(A): Electrochemical potential profile for the coefficient 
   (a,b)=$(-10^{-8},-2\times10^{-10})$ and (0,0).
   Bottom(B): Electrochemical potential profile for the coefficient 
   $(-10^{-6},-10^{-10})$. 
   Line 1st are  the electrochemical potential profile for the first order perturbative solution  
   in \eq \eqref{muu}.
   All the remaining parameters are the same as in  Fig. 2. 
 }\label{ep}
\end{center} 
\end{figure}

If we treat the system with the electric filed,  
the electrostatic potential must be  added to the chemical potential for the potential of the electron.
Now,  we consider the dimensionless electrochemical potential $\bar{\mu}=\mu-\phi$ in the specimen.
We have shown that the chemical potential and the electrostatic potential have a strong
nonlinearity for the finite current density,
but the electrochemical potential profiles decrease almost linearly for any $a$ and $b$ 
in \fig \ref{ep}(A) and \ref{ep}(B).
If the gradient of the electrochemical potential causes the electric current, 
it is reasonable that no electric current with the coefficient (0,0) shows 
the constant electrochemical potential profile in \fig \ref{ep}(A). 
Under the small external field,
the linearity of the electrochemical potential profile 
corresponds the explanation of the electron transport by Ohmic law.

These spatial dependence are reproducible by the perturbation expansion.
Now, we expand $\bar{\mu}$, $\phi$  and $\mu$ by the coefficient $a$:
$\bar{\mu}=\bar{\mu}_{0}+a\bar{\mu}_{1}+\cdots$,
$\phi=\phi_{0}+a\phi_{1}+\cdots$ and
$\mu=\mu_{0}+a\mu_{1}+\cdots$.
Substituting these $\bar{\mu}$, $\phi$ and $\mu$  into 
 \eqs \eqref{jk3}  and \eqref{EE},
we obtain the zeroth order of $a$  as
$\bar{\mu}_{0}=c_{1}$,
$\phi_{0}=c_{2}x+c_{3}$ and
$\mu_{0}=\bar{\mu}_{0}+\phi_{0}$
where the value of the each coefficient is given by 
 $c_{1}=0$, $c_{2}=-1.85\times10^{-6}$ and $c_{3}=0.0407$
to recover the profile at  $(a,b)=(0,0)$.
Expanding until the first order of $a$, 
we obtain the electrochemical potential as
\begin{equation}
\bar{\mu}=\frac{a}{c_{2}}\log(1+\frac{c_{2}}{c_{3}}x).\label{muu}
\end{equation}
Similarly, the electrostatic potential is given by
\begin{equation}
\phi=c_{2}x+c_{3}
+a\biggl[-\frac{4}{3}\frac{b}{a c_{2}^{2}}(c_{2}x+c_{3})^{3/2}+d_{1}x+d_{2}
\biggr],\label{phii}
\end{equation}
where the coefficients $d_{1}$ and $d_{2}$ are given by
\begin{equation}
d_{1}=\frac{1}{L}
\biggl[\frac{4}{3}\frac{b}{a c_{2}^{2}}
\bigl((c_{2}L+c_{3})^{3/2}-c_{3}^{3/2}\bigr)
-\frac{1}{c_{2}}\log(1+\frac{c_{2}}{c_{3}}L)
\biggr],\notag
\end{equation}
and 
\begin{equation}
d_{2}=\frac{4}{3}\frac{b}{a c_{2}^{2}}c_{3}^{3/2}.\notag
\end{equation}
The first order perturbative solution of the chemical potential  is 
given by \eqs \eqref{muu} and \eqref{phii}. 
For small electric currents, each perturbative solution plotted as the line 1st
almost correspond to  
the chemical potential profile in \fig \ref{L},
the electrostatic potential profile in \fig \ref{P}
and
the electrochemical potential  profile in \fig \ref{ep}

Now, let us estimate the gap of the chemical potential based on the approach by Nishino and Hayakawa\cite{nishino}.
We first consider the distribution function 
at the connection point.
On the left reservoir side of the connection point,
electrons  with $p_{x}\ge0$ obeying the equilibrium distribution function $f_{eq}$
go through the connection point ballistically.
In the same way, on the specimen side of the connection point,  
electrons with $p_{x}\le0$ obeying the nonequilibrium distribution function $f_{ne}$ 
go through the connection point ballistically. 
Therefore the dimensionless distribution function at the connection point is approximately given by
\begin{equation}
f_{cp}(\bm{p})=
\begin{cases}
&f_{eq}(\bm{p},\mu_{eq})\hspace{5mm}p_{x}\ge0\\
&f_{ne}(\bm{p},\mu_{ne})\hspace{5mm}p_{x}\le0,
\end{cases}\label{fcp}
\end{equation}
where the equilibrium distribution function is
\begin{equation}
f_{eq}=\frac{1}{e^{\beta(\epsilon -\mu_{eq})}+1},
\end{equation}
and the nonequilibrium distribution function as the first order of the current density is
\begin{equation}
f_{ne}=f^{0}_{ne}- \frac{3\pi^2}{2\sqrt{2}}f^{0}_{ne}(1-f^{0}_{ne})\beta v_{x}\mu_{ne}^{-3/2}j,
\end{equation}
with 
$f^{0}_{ne}=1/(e^{\beta(\epsilon -\mu_{ne})}+1)$.
The expansion of  the chemical potential 
in terms of the dimensionless current density becomes
\begin{equation}
\mu_{ne}=\mu_{eq}(1+dj)\label{d},
\end{equation}
where $d$ is the gap parameter to be determined.
Using \eq \eqref{fcp}, we define the dimensionless current density at the connection point given by
\begin{equation}
J=-2\int \frac{d\bm{p}}{(2\pi)^3}v_{x}f_{cp}.
\end{equation}
Since the electric current exists continuously at the connection point,
we assume that $J$ is equal to $j$ in \eq \eqref{jj}.
From these equations, the gap parameter is given by $d=\sqrt{3}\pi^{2}/2\mu_{eq}^2$.
Now, we consider the maximum amount of the gap at the left connection point.
We have the solution of \eq \eqref{mu},
only when the coefficient $b$  in \eq \eqref{ab} is less than $10^{-6}$.
If the specimen has the unit relative permittivity,
the maximum amount of the gap is given by $dj=-3.3\times10^{-4}$.
Therefore, we confirm that the gap of the chemical potential at the connection point is negligible. 

Now, let us dicuss our result. 
Although we are interested in the quantum effects of the electron transport,
the essence of electrons is believed to be described by semi-classical treatments
as we have discussed. 
Quantum effect may not be important for simple electron transports,
while such effects are essentially important in localization problems \cite{langer}.
We have treated the system at the isothermal low temperature, but,
it is straightforward to extend our analysis to the transport at high
temperature cases, 
where the heat conduction is as important as the electron transport.  
In such a system, the gap of the thermodynamical quantities at the connection point 
needs to be considered. The discussion of this effect is the subject of a subsequence paper.

In summary, 
we have examined the electron transport processes which is driven by the chemical potential difference
based on BGK equation coupled with Maxwell equation at the isothermal low temperature.
We determine the spatial dependence of the chemical, electrostatic and electrochemical potential as functions of the current and the impurity potential.
While the spatial dependence of the chemical potential and the electrostatic potential 
show the strong nonlinearity,
the electrochemical potential decreases almost linearly for any coefficient.
The behaviors of these thermodynamical quantities can be understood by the simple perturbation theory.
We also confirm that the gap of the chemical potential at the connection point 
between the reservoir and the specimen is negligible.

This study is partially supported by the Grants-in-Aid of Japan Space
Forum, and
Ministry of Education, Culture, Sports, Science and Technology (MEXT), 
Japan (Grant No. 18540371) and the Grant-in-Aid for the 21st century COE 
"Center for Diversity and Universality in Physics" from MEXT, Japan.

\end{document}